\documentclass[abstract=on,notitlepage,superscriptaddress,nofootinbib,aps,showpacs,twocolumn,prd]{revtex4}

\usepackage{epsf}
 \usepackage{epsfig}
\usepackage{srcltx}

\usepackage{mathrsfs}
\usepackage{amsmath}
\usepackage{amssymb}
\usepackage{graphicx}
\usepackage{dcolumn}
\usepackage{bm}
\usepackage{latexsym}
\usepackage{amssymb}
\usepackage{amsmath}
\usepackage{hyperref}
\usepackage{amsthm}
\usepackage{multirow}
\usepackage{color}

\newcommand{\be}{\begin{equation}}
\newcommand{\ee}{\end{equation}} 
\newcommand{\eei}{\end{equation}\indent\indent}
\newcommand{\bc}{\begin{center}}
\newcommand{\ec}{\end{center}}
\newcommand{\ber}{\begin{eqnarray*}}
\newcommand{\ear}{\end{eqnarray*}}
\newcommand{\ba}{\begin{array}}
\newcommand{\ea}{\end{array}}
 
\newcommand{\na}{\nabla}

\newcommand{\ti}{\tilde}

\newcommand{\bea}{\begin{eqnarray}}
\newcommand{\eea}{\end{eqnarray}}
\newcommand{\nn}{\nonumber}

\newcommand{\ei}{\end{itemize}}

\newcommand{\bra}[1]{\left(#1\right)}
\newcommand{\bras}[1]{\left[#1\right]}

\newcommand \veps {\varepsilon} 

\newcommand{\Lietwo}{{\cal L}}

\newcommand{\reff}[1]{(\ref{#1})}

\def\case#1/#2{\textstyle\frac{#1}{#2} }

\begin{document}

 \preprint{}
\title{Echoes from the black holes: Evidence of higher order corrections to General Relativity in strong gravity regime}
\author{Dan B. Sibandze}
\email{danx36@gmail.com}
\author{Rituparno Goswami}
\email{goswami@ukzn.ac.za}
\author{Sunil D. Maharaj}
\email{maharajs@ukzn.ac.za}
\affiliation{ Astrophysics and Cosmology Research Unit, School of Mathematics, Statistics and Computer Science, University of KwaZulu-Natal, Private Bag X54001, Durban 4000, South Africa.}
\author{Peter K. S. Dunsby}
\email{peter.dunsby@uct.ac.za}
\affiliation{Department of Mathematics and Applied Mathematics and ACGC, University of Cape Town, Cape Town,7701,  South Africa.}

\begin{abstract}
We show that the higher order curvature corrections to general relativity in the strong gravity regime of near horizon scales produce a rapidly oscillating and infalling Ricci scalar fireball just outside the horizon. This can generate the ringdown modes of gravitational waves having the same natural frequency as those which are generated by the black hole mergers. Our analysis provides a viable explanation to the echoes in the ringdown modes recently detected from the LIGO data, without invoking the existence of any exotic structures at the horizon.
\end{abstract}


\maketitle
\section{Introduction}

The detection of gravitational waves from binary black hole mergers \cite{Ligo1,Ligo2,Ligo3} was a historic event that put General Relativity (GR) on an even stronger footing as the most favored classical theory of gravitational interactions. However, the key question still remains: Is GR an effective theory in the weak field regime of some more general theory? This question is inspired by the currently available non-perturbative quantum theories of gravity like String Theory and Loop Quantum Gravity, where we expect significant deviations from the classical results on scales of the order of the Black Hole (BH) horizon. More importantly, if indeed such non-GR features are present near BH horizons, there should be signatures of these in the gravitational waves emitted from the mergers. These signatures cannot be explained by GR.

It is therefore extremely interesting that preliminary investigations of the LIGO open data \cite{Ligo4,Ligo5,Ligo6} revealed the existence of {\em echoes} of the ringdown modes \cite{echoe}, which would not be possible without the presence of some semiclassical structures near the horizon.  By building a phenomenological template for successive echoes from exotic quantum structures expected in firewall or fuzzball models or exotic compact objects (ECO's) \cite{eco}, and after marginalizing over its parameters, tentative evidence for these echoes were reported at $2.9\sigma$ significance. Although these echo templates are somewhat {\it adhoc} in nature, it is expected that future numerical simulations of merging black holes, where the horizon is replaced by a membrane structure can sharpen the echo template.

In this paper, we clearly show that if we consider higher order curvature corrections to the general relativistic Lagrangian on scales of the order of the horizon, a {\it fireball} of very high frequency fluctuations of Ricci scalar will be produced near the horizon. These fluctuations behaves like an infalling extra massive scalar field that can produce the echoes of the ringdown modes as detected in the LIGO data. It is important to note here that these higher order corrections are not {\it adhoc} in nature, but would be expected from any attempt to create a renormalisable theory of gravity (like string theory). The advantage of this approach is two fold: firstly this result indicates the existence of a more general theory of gravity in the strong gravity regime, of which GR is an weak field approximation. Secondly we do not need to invoke any exotic objects to explain the echoing effects during a black hole merger. 

\section{Higher order curvature corrections to general relativity}

In GR, the Einstein-Hilbert Lagrangian density of gravitational interaction is given by
\be
 \Lietwo_{EH} = \sqrt{-g} \bra{R-2\Lambda}\,.
 \ee
We can generalise the above Lagrangian density by adding the higher order curvature correction terms generated by the Riemann curvature tensor:
\bea
\Lietwo_{g} &= &\sqrt{-g}\left[R-2\Lambda+\alpha R^2 + \beta R_{ab}R^{ab}+\gamma R_{abcd}R^{abcd}\right.\nn\\
&&\left.+\nu \veps^{klmn} R_{klst}R^{st}{}_{mn}+\cdots+\mathcal{O}(N)\right],
 \eea
where $\alpha,\beta,\gamma$ and $\nu$ are coupling constants. In fact, as discussed earlier, in the quantum field picture the effects 
of renormalisation are expected to add such terms to the Lagrangian in order to give a first approximation to some quantised theory of gravity \cite{DeWitt:1967,Birrell+Davies}. 
Retaining up to the quadratic terms and using the very well known results \cite{DeWitt:1965,Buchdahl:1970,Barth:1983}, 
\begin{align}
(\delta/\delta g_{mn})\int &  dV \bra{R_{abcd}\,R^{abcd}-4R_{ab}\,R^{ab}+R^2}=0\;,\\
(\delta/\delta g_{ab})\int & dV \,\veps^{klmn} R_{klst}\,R^{st}{}_{mn}= 0\;,
\end{align}
we can eliminate the Kretchman scalar term and the term involving $\epsilon^{iklm}R_{ikst}\,R^{st}{}_{lm}$ from the Lagrangian density. Also it has been shown that the theories that contain the square of the Ricci tensor in the action, suffer from several instabilities like the Ostrogradsky instability \cite{Ostrogradsky:1850}. Therefore the Lagrangian density, up to quadratic order, of a stable gravitational theory will only contain the square of the Ricci scalar and the corresponding gravitational action can be written as 
\be
{\cal S} =\frac12 \int dV\sqrt{-g} \bra{R+ \alpha R^{2}}\,.
\label{generalfr}
\ee
\subsection{Constraints on the coupling constant}

As discussed in detail in \cite{3GR}, Solar system experiments as well as cosmological observations give a strong bound on the coupling constant $\alpha$. Perhaps the strongest constraint is given by the latest data from Planck, which does not rule out the above extended  gravitational theory (\ref{generalfr}) as a viable candidate for the early acceleration phase of the universe. However for this to be indeed the case we must have  \cite{Starobinsky:2007hu, Starobinsky:1983zz} $\alpha\simeq 10^{-45}\bigl(N/50\Bigr)^2 \text{eV}^{-2}$ where $N$ is the number of e-folds. Hence for all practical purposes we will consider $\alpha\simeq 10^{-45} \text{eV}^{-2}$ from the cosmological constraints. It is evident that for such a small value of coupling constant GR remains the best fit theory in the weak gravity regime.
\subsection{Curvature corrected field equations in vacuum}
Varying the action (\ref{generalfr})  with respect to the metric $g_{ab}$ yields:
\bea
\delta{\cal S} &=& -(1/2) \int dV  \, \sqrt{-g} \left\{(1/2)\bra{R+ \alpha R^{2}}( \, g_{ab} \, \delta g^{ab}\right. \nonumber\\
&&\left. -\bra{1+ 2\alpha R} \,\delta R \,  \right\} ~.
\eea
Since $R= g^{ab}\,R_{ab}$ and the connection is the Levi-Civita one, we can write
\bea
 \bra{1+ 2\alpha R}  \,\delta R &\simeq  &\delta g^{ab}\left[\bra{1+ 2\alpha R} \, R_{ab}+ 2\alpha g_{ab} \, \Box R\right. \nn \\
 &&\left. - 2\alpha\na_{a}\na_{b}R\right],
\eea
where the $\simeq$ sign denotes equality up to surface terms and 
$\Box \equiv \na_{c}\na^{c}$. By requiring that $\delta{\cal S} =0$ with respect to variations in the metric, we finally get the required field equation:
\bea
 \bra{1+ 2\alpha R} G_{ab}&= &- \frac12 g_{ab}\,\alpha R^2+2\alpha\na_{a}\na_{b}R\nonumber\\
 &&- 2\alpha g_{ab} \,\Box R\, .
  \label{field1} 
\eea
Here $G_{ab}$ is the Einstein tensor, and we can easily see that when $\alpha=0$, we regain the Einstein field equations in vacuum. Taking the trace of the field equations above, we get 
\be\label{trace}
6\alpha\Box R-R=0\;.
\ee
This is a non-trivial equation that determines the evolution of Ricci scalar in vacuum.
\subsection{Comparison with GR}

We would now like to highlight the key similarities and differences from GR, when we consider the curvature corrected field equations in vacuum.
\begin{itemize}
\item {\bf Similarities}: From the field equations (\ref{field1}) it is evident that all the Ricci flat ($R=0$) vacuum solutions of GR are a solution of the curvature corrected theory. This implies that at the level of the background, the Schwarzschild or Kerr geometries remain a solution to this theory. Since these geometries encompass all the possible astrophysical black hole spacetimes, it follows that there will be absolutely no difference in the properties of the black holes at the background level.
\item {\bf Differences}: The key difference arises when we consider small perturbations around these background geometries. In GR we know that the Ricci scalar has to vanish in vacuum. Hence any small geometrical perturbations of the background geometry will not affect the Ricci scalar. However in the curvature corrected theory, because of the non-trivial trace equation (\ref{trace}), we see that there can be small perturbations of the Ricci scalar around it's zero value in the background. This will then generate a Ricci scalar wave together with the usual tensor gravitational wave degrees of freedom.
\end{itemize}
\section{Ricci Wave fireball around perturbed black holes}
For a more detailed analysis of the Ricci wave phenomenon, let us consider a Schwarzschild black hole perturbed from it's usual background geometry (as one would expect just after the black hole merger). This will then perturb the Ricci scalar from it's zero background value and it's evolution will be governed by the trace equation (\ref{trace}). Seeking the solution of this equation of the form $R(r,t)\equiv e^{i\kappa t}R(r)$, and performing the usual harmonic decomposition for the d'Alembert operator in the Schwarzschild geometry and using the tortoise coordinates $r*$, the trace equation takes the form \cite{anne}
\begin{equation}
\left(\frac{d^2}{dr^{2}_{*}}+\kappa^{2}-V_{S}\right)\mathcal{R}=0 \label{ode1}
\end{equation}
where we have rescaled $R=r^{-1}\mathcal{R}$, and 
\begin{equation}\label{Scalpot}
V_{S}=\left(1-\frac{2m}{r}\right)\left[\frac{l(l+1)}{r^2}+\frac{2m}{r^3}+\frac{1}{6\alpha}\right]\;,
\end{equation}
is the Regge-Wheeler potential for the Ricci scalar perturbations, with $m$ being the black hole mass.  The form of the wave equations \reff{ode1} is similar to a one dimensional Schr\"{o}dinger equation and hence the potential corresponds to a single potential barrier. This equation can be made dimensionless by multiplying through with the square of the black hole mass $m$. In this way the potential \reff{Scalpot} 
becomes
\bea
V_S&=&\bra{1-\frac{2}{r}}\bras{\frac{\ell\bra{\ell+1}}{r^2}
+\frac{2}{r^3}+\frac{1}{6\alpha}}~,
\label{normalpotential}
\eea
where we have defined (and dropped the tildes)
\be
\tilde{r} = \frac{r}{m}~, \qquad \ti{\alpha} = \frac{\alpha}{m}~, \qquad \ti{\kappa}=m\kappa \;.
\label{dimensionless}
\ee

It is interesting to note that the equation (\ref{ode1}) is exactly the same as a massive scalar wave equation in Schwarzschild background with the scalar field mass $\mathcal{M}=\sqrt{\frac{1}{6\alpha}}$. Now let us look the property of this equation carefully. At the horizon ($r_{*}\rightarrow -\infty$),  we have $V_{S}=0$, and equation (\ref{ode1}) becomes
\begin{equation}
\left(\frac{d^2}{dr^{2}_{*}}+\kappa^{2}\right)\mathcal{R}=0\;, \label{ode2}
\end{equation}
with two linearly independent solutions
\be\label{sol1}
\mathcal{R}\sim C_1 \exp{(i\kappa r_{*})} + C_2 \exp{(-i\kappa r_{*})}\;.
\ee
Since there cannot be any outgoing modes at the horizon, this implies $C_2 = 0$.
At spatial infinity ($r_{*} = +\infty$), equation (\ref{ode1}) becomes
\begin{equation}
\left(\frac{d^2}{dr^{2}_{*}}+\kappa^{2}-\mathcal{M}^2\right)\mathcal{R}=0\,. \label{ode2}
\end{equation}
The solution for ingoing modes is given as 
  \be\label{infty}
 \mathcal{R}\sim C_3\exp{(i\sqrt{\kappa^2-\mathcal{M}^2}r_*)} \,.   
 \ee

Now from the previous section we know that $\alpha<<1$, which means  $\mathcal{M}>>1$. Hence this problem reduces to the problem of an infalling massive scalar field into the black hole. Now for all frequencies $\kappa<\mathcal{M}$, from equation (\ref{infty}) we can immediately see that the solution goes to zero exponentially at the spatial infinity. 
We can now solve the Ricci wave equation (\ref{ode1}) numerically, for a realistic black hole with $\mathcal{M} >> k$, using the following boundary conditions:
\be
 \mathcal{R}\sim 0 \quad at \quad (r_{*} = +\infty)
  \ee
and
\be
 \mathcal{R}\sim e^{i\kappa r_*}  \quad at \quad (r_{*} = -\infty)\;,
  \ee
In figure 1 we have plotted the nature of the Ricci scalar perturbations around the black hole. It has an interesting behaviour: as $\mathcal{M}$ increases the Ricci scalar fluctuates with extremely high frequency near the horizon and rapidly dies down to zero value within  $r_{*} = 0 \Rightarrow r \sim 2.5$. Thus we can conclude that a perturbed black hole in a curvature corrected theory is surrounded by a rapidly oscillating and infalling Ricci scalar field just outside the horizon. Thus without invoking any quantum phenomenon we can get a massive scalar {\it fireball} surrounding the black hole horizon.
\begin{figure}[ht] 
\caption{Ricci scalar perturbations around the black hole for $k=5$, $\mathcal{M}=500$}
   \centering
   \includegraphics[width=6cm]{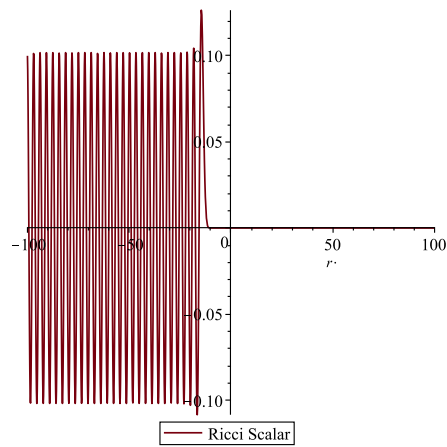} 
\end{figure}
 
 \section{Quasinormal modes due to massive scalar accretion}
 In the last section we established that a higher order correction to GR on scales of the order of the horizon gives rise to rapidly oscillating Ricci scalar just outside the horizon and it behaves exactly like infalling massive scalar field. Now our problem reduces to the following: {\it We have a Schwarzschild black hole with an infalling massive scalar test field in the exterior Schwarzschild geometry. We would like to know the nature of gravitational waves produced by the black hole which is perturbed due the the presence of this accreting massive test field.} Fortunately a very detailed analysis of the above problem has already been performed in \cite{scalar1}, which generalized all the important earlier works \cite{scalar2,scalar3}.
 The key findings of these papers are as follows:
 \begin{enumerate}
 \item The gravitational wave generated by the infall of the massive scalar field has some unique features that differentiates it from those generated  
black hole mergers or by the infall of dust. The most interesting feature is that the ring-down part of the gravitational wave in case of massive scalar accretion has the same values of the quasinormal frequencies as those obtained in the case of a binary black hole collision. Hence just from the ring down part it is very hard to differentiate between these processes.
\item The above point is really interesting as it shows that although the frequency of the scalar field propagating on a Schwarzschild background is
different from the one associated with the gravitational perturbation, the gravitational signal preserves its characteristic ring-down frequency. This happens despite the fact that the
scalar wave travels together with the gravitational one.
\item The late time tails of the gravitational waves generated due to the infalling massive scalar field do differ from that of a binary black hole merger and this gives a nice observational test for differentiating these processes. 
\item The amplitude of the emitted gravitational waves due to the massive scalar accretion increases as the mass increases. 
\end{enumerate}
We can therefore safely claim that the infalling Ricci waves due to the curvature corrected theory will generate gravitational waves with the same natural frequency as the binary black hole merger and that is exactly what causes the echoes in the ringdown modes.

\section{Conclusions - Putting it all together}

In this paper we propose a viable explanation for the {\it echoes} of the ringdown modes from the binary black hole mergers detected by LIGO, without invoking the existance of any exotic structures near the black hole horizon. Inspired by the available renormalisable quantum theories of gravity, we conjecture that there should be higher order curvature corrections to GR on scales of the order of the horizon. We showed that these corrections produce rapidly oscillating and infalling Ricci scalar waves near the horizon that behaves exactly like an accreting massive scalar field. As already known, the perturbed black holes due to this massive scalar accretion produces gravitational waves that have exactly the same natural characteristic ringdown frequency as those generated by the binary black hole mergers. It is exactly these waves that are detected as the {\it echoes from the abyss.}

\begin{acknowledgments}
All the authors are supported by National Research Foundation (NRF), South Africa. SDM 
acknowledges that this work is based on research supported by the South African Research Chair Initiative of the Department of
Science and Technology and the National Research Foundation.
\end{acknowledgments}


\end{document}